# The Diversity Multiplexing Tradeoff for Interference Networks

Aydin Sezgin, Syed Ali Jafar, and Hamid Jafarkhani[1]
University of California, Irvine, USA
email:{asezgin,syed,hamidj}@uci.edu

*Abstract*— The diversity-multiplexing tradeoff (DMT) for interference networks, such as the interference channel, the X channel, the Z interference channel and the Z channel, is analyzed. In particular, we investigate the impact of rate-splitting and channel knowledge at the transmitters. We also use the DMT of the Z channel and the Z interference channel to distill insights into the "loud neighbor" problem for femto-cell networks.

## I. INTRODUCTION

The tradeoff between rate and reliability of a system is captured asymptotically in the high SNR regime by the *diversity-multiplexing tradeoff* (DMT) [1]. This tradeoff is especially relevant for wireless networks where the time scale of dynamic behavior is too long for ergodic capacity to be a meaningful metric. The DMT is characterized for the multiple access channel (MAC) in [2]. While allowing different multiplexing gains (a.k.a. degrees-of-freedom (DoF) ) for each user, [2] assumes equal diversity requirements and equal number of antennas at each user. The result is generalized in [3] where each user is allowed to have different diversity and multiplexing gain requirements as well as a different number of antennas. While the conventional DMT for the MIMO broadcast channel (BC) has not been explicitly reported, a related multiuser-diversity versus spatial multiplexing tradeoff is studied in [4]. The DMT for the 2-user interference channel with single-antenna nodes is explored in [5] and extensions to multiple-antenna nodes appear in [6]. The DMT for the $Z$ interference channel is studied in [7] and the results are extended to linear ad-hoc networks in [8].

An important issue that is occasionally overlooked in some of these works is the role of rate-splitting in interference networks. Splitting a user's message into common and private messages is a key feature of the Han-Kobayashi [9] achievable scheme for the 2 user interference channel. The private messages are decoded only by the desired users while the common messages are decoded by both users. Decoding a part of the interference (corresponding to the common message) leads to partial interference cancelation and is the key to the capacity characterization within 1 bit found in [10]. The achievable scheme of [10] uses Gaussian codebooks with superposition coding and sets the power of the private message so that it is received at the noise-floor level at the unintended receiver. Channel knowledge is needed at the transmitter, most notably in order to set the private message power at the noise-floor level of the undesired receiver. The role of channel uncertainty at the transmitters is explored in [11] through the compound interference channel model. In this model, the channel coefficients can take any one out of many possible values and the channel realization, unknown to the transmitters, is held fixed throughout the duration of the communication. For this case, [11] finds that an approximately optimal (within one bit of capacity outerbound) achievable scheme splits the message into multiple parts, and the receivers opportunistically decode as much of the interference as their channel realization allows. We point out some of the issues regarding rate-splitting in prior works that we resolve in this paper.

The possibility of rate-splitting is not considered in the definition of the outage event in [7], [8] where the only decoding possibilities considered correspond to decoding all or none of the interfering signal. We show that in general this outage event definition is sub-optimal. However, for the purpose of the asymptotic DMT characterization we validate the assumptions of [7], [8]. In fact we show that, for the purpose of DMT, not only is it enough to ignore rate-splitting, but it suffices to consider only the achievable scheme where all messages are to be decoded by all decoders.

A different issue arises with [5], [6] which claims that the rate region whose achievability is known only with rate-splitting and channel knowledge at the transmitters (CSIT), is achievable even without channel knowledge. In the MIMO scenario, it is claimed that if the overall transmit covariance matrix is chosen to be the scaled identity matrix, then the rate region given by the Chong-Motani-Garg [12] simplification of the Han-Kobayashi [9] achievable region, is achievable without CSIT. The problem with this claim is that the Chong-Motani-Garg achievable region, in its compact form (where the specific rate allocations for private and common messages are eliminated by Fourier-Motzkin elimination), implicitly uses channel knowledge in deciding the rate allocation to private and common messages, even if the power allocations are chosen blindly. We explicitly characterize the interference channel DMT with no CSIT and find that it is strictly smaller than the DMT claimed in [5], [6], which in fact correspond to the DMT with full CSIT.

Lastly, in this work we focus on the special case of the Z channel and the Z interference channel - which conceptually model the interference problem encountered in femto-cell networks, identified as the "loud neighbors" problem in [13].

[1]This work was supported by NSF CAREER Grant CCF-0546860 and CCF-0930809 and by DARPA ITMANET under Grant UTA06-793.

The DMT of the Z channel and the Z-interference channel are used to distill comparative insights into the "loud neighbor" problem for femto-cell networks under open-access, closed-access and orthogonal access models.

## II. SYSTEM MODEL

Consider a $S$ user X network where independent messages $W_{sd}$, $1 \leq s \leq S$, $1 \leq d \leq D$, need to be communicated from transmitter $s$ to receiver $d$ respectively. Here, we have $S = D$. Note that the X network includes multiple access, broadcast and interference channels as special cases. Over $n$ channel uses, the system input-output equations are the following:

$$Y_d^n = \sum_{s=1}^{S} H_{sd} X_s^n + Z_d^n,$$

where $Y_d^n$ is the vector of received symbols at destination node $d \in S$, $X_s^n$ is the vector of transmitted symbols from source $s \in S$ and is a function of the message originating at that source, i.e. $X_s^n(W_s)$. $Z_d^n$ is the additive white Gaussian noise sequence, and $H_{sd}$ is the channel between source $s$ and destination $d$. The channel is held fixed throughout the duration of the communication. Further, we assume throughout that the channel values are generated independent, identically and $\mathcal{CN}(0, \sigma_{sd}^2)$ distributed, with $\sigma_{sd}^2 = 1$, although some of our results apply to more general distributions. The channel realization is known to all receivers but not to the transmitters. The receivers employ the optimal (maximum a-posteriori probability) decoding rule, and $\hat{W}_{sd}(Y_d^n)$ are the decoded messages. The transmit power constraint is expressed as $E[|X_s|^2] \leq \mathsf{SNR}$, where $\mathsf{SNR}$ is the signal-to-noise-ratio. Note that we are considering large block lengths, such that outage is the dominant error event rather than noise. The average probability of error for source $s$ at destination $d$ is defined as the probability that the decoded message $\hat{W}_{sd}$ is not equal to the transmitted message $W_{sd}$ averaged over the equally likely messages and the channel realizations, i.e.

$$P_e^{(sd)} = \text{Prob}\left[\hat{W}_{sd}(Y_d^n) \neq W_{sd}\right].$$

The overall probability of error is

$$P_e = \text{Prob}\left[\bigcup_{s,d} \hat{W}_{sd}(Y_d^n) \neq W_{sd}\right] \doteq \max_{s,d} P_e^{(s,d)},$$

i.e. we consider the overall system to be in outage if there is at least one user in outage. The spatial multiplexing or degree-of-freedom (DoF) $r_{sd}$ for the link between source $s$ and destination $d$ is defined as

$$r_{sd} = \lim_{\mathsf{SNR} \to \infty} \frac{R_{sd}(\mathsf{SNR})}{\log \mathsf{SNR}},$$

where $\log \mathsf{SNR}$ is the capacity of an additive white Gaussian noise (AWGN) channel at high SNR. The diversity gain $d_{sd}$ is defined as

$$d_{sd}(\mathbf{r}) = -\lim_{\mathsf{SNR} \to \infty} \frac{\log P_e(\mathbf{r})}{\log \mathsf{SNR}}.$$

where $\mathbf{r} = (r_{11}, \ldots, r_{sd}, \ldots, r_{SD})$. We define the overall diversity gain as

$$d(\mathbf{r}) = \min(d_{11}(\mathbf{r}), \ldots, d_{sd}(\mathbf{r}), \ldots, d_{SD}(\mathbf{r})). \quad (1)$$

In the following, different assumptions are made regarding the availability of CSIT. While outage is well defined in case the transmitters are not aware of the instantaneous channel realizations, it is not clear immediately what that means if the transmitters have channel knowledge and thus a short discussion might be in order. Since the transmitters are aware of the current channel status, they know in advance which rates can be reliably achieved. If the channel quality falls below a critical level certain rates cannot be guaranteed and the transmitter declares a failure and stops transmitting until the channel changes in a block fading manner. Thus, with a probability $Pr$ certain rates are guaranteed to be achievable and the DMT with CSIT characterizes the slope (diversity gain) of failure probabilities as a function of the rates or multiplexing gains. In the following section, we characterize the DMT without CSIT, while Section IV is devoted to the case with CSIT.

## III. DMT WITHOUT CSIT

### A. DMT of the X Channel

First, the DMT for the multiple access channel (MAC) with asymmetric configurations is stated, which is needed in the following [1].

*Theorem 1 ([3]):* Consider a MAC with $S$ users, where the $s$th user has $M_s$ transmit antennas and the base station has $N$ receive antennas. The optimal diversity-multiplexing tradeoff for any user $s$ in the MAC, given an achievable rate vector, $(r_1, \ldots, r_S)$, is described by

$$d_s^{MAC}(r_1 \ldots, r_S) = \min_{\mathcal{S}_s} d_{\sum_{s' \in \mathcal{S}_s} M_{s'}, N}\left(\sum_{s' \in \mathcal{S}_s} r_{s'}\right)$$

where $\mathcal{S}_s = \{\{s\} \cup \tilde{\mathcal{S}}_s, \forall \tilde{\mathcal{S}}_s \subseteq \{1, \ldots s-1, s+1, \ldots, S\}\}$ and $d_{M,N}$ is the DMT of the point-to-point MIMO channel given by $d_{M,N}(r) = (M-r)^+(N-r)^+$, where $(x)^+ = \max(0, x)$. Thus, the overall diversity gain is given as $d^{MAC}(r_1 \ldots, r_S) = \min_s d_s^{MAC}$. The DMT of the $S$ user X network [14] is obtained in the following theorem.

*Theorem 2:* The DMT of the $S$ user X network, where the transmitter $s$ is equipped with $M_s$ antennas and the receivers are equipped with $N$ antennas is the following

$$d^X(\mathbf{r}) = d^{MAC}(r_1, \ldots, r_S)$$

with $r_s = \sum_{d=1}^{D} r_{sd}$.

*Proof:* Since the outputs are statistically equivalent, i.e.,

$$p(Y_d^n | X_1^n, \ldots X_S^n) = p(Y_{\tilde{d}}^n | X_1^n, \ldots X_S^n)$$

with $d \neq \tilde{d}$, it follows

$$\Pr(\hat{W}_{sd}(Y_d^n) \neq W_{sd}) = \Pr(\hat{W}_{sd}(Y_1^n) \neq W_{sd})$$

---
[1] In this section, we implicitly assume that all channels have equal strength.

with $d \neq 1$, the probability of error of each message at its desired receiver is the same as its probability of error at receiver 1. Therefore the DMT of the X network is the same as the DMT of the MAC at receiver 1. ∎

As a special case, we have the following corollary.

*Corollary 1:* The DMT of the 2 user interference channel is the following: $d^{IC}(r_1, r_2) = d^{MAC}(r_1, r_2)$.
In other words, the DMT of the 2 user interference channel with all single antenna nodes, is the same as the DMT of a 2 user multiple access channel with all single antenna nodes.

In particular when $r_1 = r_2 = r$, the DMT can be expressed as: $d^{IC}(r,r) = \min((1-r)^+, 2(1-2r)^+)$, which is different from the expression obtained in [5] $d^{IC}(r,r) = \min((1-r)^+, 3(1-2r)^+)$. Indeed, the latter expression is in fact the DMT for the 2 user interference channel with full CSIT. As noted in the introduction, a key feature of the Han-Kobayashi region [9] (or its compact version, the Chong-Motani-Garg region [12]) is that each source splits its message into two parts, the private and the common information. At each destination, the private information of the interfering source is treated as noise, while the common information of the interfering source, the common and the private information of the intended source are decoded simultaneously. The achievable rate region is only obtained, if the rates and powers allocated to the common and the private part are performed according to [12, Theorem 1]. If the sources are not aware of the channel state, then we have shown that the Han-Kobayashi region is not achievable, since the DMT region without CSIT is strictly contained inside the DMT with CSIT. This is claimed differently in [5] in the single-antenna case. In the multi-antenna case, using an argument similar to that developed in [1], it is claimed that using a scaled identity matrix the simplified Han-Kobayashi-Region used in [10] with fixed power splitting and without time-sharing is achievable even without CSIT.

As a final remark, the result for the BC is obtained from Theorem 2 by setting $r_{sd} = 0$ for $s \neq 1$. It follows that for the SISO BC, without CSIT a reasonable strategy is to apply time-sharing and transmit to the users in a consecutive fashion. Thus the DMT is similar to the point-to-point channel. For the MIMO BC with $M$ transmit antennas at the source and $N$ antennas at the users, again the outputs at the users are statistically equivalent. Assuming symmetric rate requirements $r$, using TDMA seems to be a reasonable strategy.

In the following section, we characterize the DMT for the Z interference channel with two users.

*B. DMT of the Z interference channel*

The Z interference channel is a special case of the interference channel where $H_{12} = 0$, i.e. there is no interference from transmitter 1 to receiver 2. The DMT of the Z interference channel is explored in [7], where the decoding at receiver 1 is restricted to two possibilities - either both messages $W_1, W_2$ are decoded jointly as in a multiple access channel, or the signal for message $W_2$ is treated as noise. Thus, the setup ignores the possibility of rate-splitting and partial interference cancelation. As shown by [11], rate-splitting and partial interference cancelation can in general reduce the probability of outage. The numerical example in Fig. 1 illustrates this point at finite SNR. From Fig. 1, we observe that using rate-splitting in this example gives a gain of about 2 dB at an outage probability of $6 \cdot 10^{-2}$. Note that the rate and power split between the private and common message is not optimized for the rate-splitting curve in Fig.1. The change of slope of the curve is due to the effect that up to a certain SNR the outage probability is dominated by the strategy of treating interference as noise. For higher SNR it turns out to be better to decode (part of) the interfering signal. However, the following theorem shows that rate-splitting is not useful for the DMT of the Z interference channel. In fact, we show that even the possibility of treating the interfering signal as noise is not useful for the DMT of the Z interference channel and receiver 1 may be constrained to decode both $W_1, W_2$ without loss in DMT.

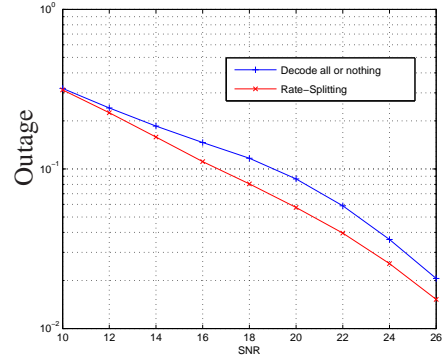

Fig. 1. Outage performance with and without rate-splitting for the Z-interference channel. The target rates were set to $(R_{11}, R_{21}, R_{22}) = (2, 0.4, 2)$, channel variances given as $(\sigma_{11}^2, \sigma_{21}^2, \sigma_{22}^2) = (1, 0.03, 1.4)$ and the power split at the second user was 0.7 (common) to 0.3 (private).

*Theorem 3:* The DMT of the Z interference channel is given by
$$d^{ZIC}(r_1, r_2) = d^{MAC}(r_1, r_2).$$

*Proof:* Consider any coding scheme for the Z interference channel that achieves multiplexing gain $(r_1, r_2)$ and diversity gain $d^{ZIC}(r_1, r_2)$, so that:
$$\Pr[\hat{W}_1(Y_1^n) \neq W_1] \leq \mathsf{SNR}^{-d^{ZIC}(r_1,r_2)+o(1)}$$
$$\Pr[\hat{W}_2(Y_2^n) \neq W_2] \leq \mathsf{SNR}^{-d^{ZIC}(r_1,r_2)+o(1)}$$

Now, with the same coding scheme consider the multiple access channel at receiver 1. For the multiple-access channel, we can write
$$\begin{aligned}P_e =&\Pr[\hat{W}_1(Y_1^n) \neq W_1 \text{ OR } \hat{W}_2(Y_1^n) \neq W_2]\\=&\Pr[\hat{W}_1(Y_1^n) \neq W_1]\\&+\Pr[\hat{W}_2(Y_1^n) \neq W_2|\hat{W}_1(Y_1^n) = W_1]\Pr[\hat{W}_1(Y_1^n) = W_1]\\=&\Pr[\hat{W}_1(Y_1^n) \neq W_1]\\&+\Pr[\hat{W}_2(Y_2^n) \neq W_2]\Pr[\hat{W}_1(Y_1^n) = W_1]\\\leq&\mathsf{SNR}^{-d^{ZIC}(r_1,r_2)+o(1)}.\end{aligned}$$

It follows that

$$d^{MAC}(r_1, r_2) \geq d^{ZIC}(r_1, r_2).$$

But, since the Z interference channel is obtained from the interference channel by removing one interfering link,

$$d^{ZIC}(r_1, r_2) \geq d^{IC}(r_1, r_2).$$

Since $d^{IC}(r_1, r_2) = d^{MAC}(r_1, r_2)$ from Corollary 1, we have

$$d^{MAC}(r_1, r_2) \geq d^{ZIC}(r_1, r_2) \geq d^{MAC}(r_1, r_2).$$

∎

In the following section, we characterize the DoF and the DMT of the Z channel and Z interference channel with perfect channel knowledge at the receivers to distill comparative insights into the "loud neighbor" problem for femto-cell networks under open-access, closed-access and orthogonal access policies.

## IV. LOUD NEIGHBORS IN FEMTO-CELLS - A DMT PERSPECTIVE WITH CSIT

The deployment of femto-cells is one way to improve the coverage in particular for residential or small business environments, i.e. in areas with limited or unavailable access to a base station. A femto-cell is a small cellular base-station, which connects to the network of the wireless service provider using a broadband connection as the backhaul. Subscribers to the femto-cell in those residential environments benefit in terms of battery life as well as link quality. However, nearby mobile stations, which are referred to as macro-cell users, not associated to the femto-cell still suffer from the lack of mobile network coverage. These macro-cell users usually try to improve the connectivity by transmitting with increased power. As a consequence of this "loud neighbor" effect, the femto-cell might suffer from severe interference. It is therefore of interest to analyze the performance of networks with femto-cells using different policies on how to deal with macro-cell users. First of all, the system equations for this specific setup reduce to $Y_f^n = H_{11}X_1^n + H_{21}X_2^n + Z_1^n$ and $Y_m^n = H_{22}X_2^n + Z_2^n$, where $Y_f^n$ denotes the received signal at the access point of the femto-cell and $Y_m^n$ denotes the received signal at the BS. The signal-to-noise ratio and the interference-to-noise ratio of the femto-cell are $\mathsf{SNR}_f$ and $\mathsf{INR}_f$, respectively. The signal-to-noise ratio of the direct link from the macro-cell user to the BS is $\mathsf{SNR}_m$.

As aforementioned, we use the Z channel as a model for the femto-cell "loud neighbor" problem. Models for femto-cell policies include the closed-access, open- access, and the orthogonal access models. In the closed-access model, the macro-cell user is not allowed to use the femto-cell access point to send his data. The information theoretic model for this scenario is the Z interference channel. An outer bound on the capacity region of the Z-interference channel is obtained by the outer bound on the capacity region of the interference channel in [5] by setting $H_{12} = 0$. Further, we define the ratio $\alpha = \log \mathsf{SNR}_m / \log \mathsf{SNR}_f$. Using the upper bound on the capacity region from [5] with $H_{12} = 0$ and the definitions above, in the high SNR limit we have

$$r_{11} \leq 1, \quad r_{22} \leq \alpha, \quad r_{11} + r_{22} \leq \max(1, \alpha).$$

In the following proposition, we characterize the DMT of the closed-access policy. The proof of the proposition follows the same line of arguments as in [5] and is omitted due to the lack of space.

*Proposition 1:* The diversity order of the system with closed-access is given by $d_{out}(r_{11}, r_{22}, \alpha) = \min(d_a, d_b, d_c)$, where

$$d_a(r_{11}) = (1 - r_{11})^+, \quad d_b(r_{22}, \alpha) = (\alpha - r_{22})^+$$
$$d_c(r_\Sigma, \alpha) = 2(1 - 2r_\Sigma)^+ + (\alpha - 2r_\Sigma)^+$$

with $r_\Sigma = r_{11} + r_{22}$.

In the open-access model, the macro-cell user is allowed to use the femto-cell access point to send his data. The information theoretic model for this scenario is the Z channel with three messages. An upper bound to the capacity of the Z channel is given by (obtained by using the outer bounds of the X-channel in [14] with $H_{12} = 0$, $R_{12} = 0$, and $R_2 = R_{21} + R_{22}$)

$$\mathcal{R}_0(H_{11}, H_{21}, H_{22}) = \Big\{ (R_{11}, R_2) \in \mathbb{R}_+^2 :$$
$$(a) \quad R_{11} \leq \log\left(1 + \mathsf{SNR}_f |H_{11}|^2\right)$$
$$(b) \quad R_2 \leq \log\left(1 + \max\left(\mathsf{SNR}_m|H_{22}|^2, \mathsf{INR}_f|H_{21}|^2\right)\right)$$
$$(c) R_{11} + R_2 \leq \log\left(1 + \mathsf{SNR}_f|H_{11}|^2 + \mathsf{INR}_f|H_{21}|^2\right)$$
$$+ \log\left(1 + \frac{\mathsf{SNR}_m|H_{22}|^2}{1 + \mathsf{INR}_f|H_{21}|^2}\right) \quad (2)$$
$$(d) R_{11} + R_2 \leq \log\left(1 + \mathsf{SNR}_f|H_{11}|^2 + \frac{\mathsf{INR}_f|H_{21}|^2}{1 + \mathsf{SNR}_m|H_{22}|^2}\right)$$
$$+ \log\left(1 + \mathsf{SNR}_m|H_{22}|^2\right) \Big\}.$$

Using this upper bound on the capacity region for the Z channel, in the high SNR limit we have

$$r_{11} \leq 1, \quad r_2 \leq \max(\alpha, 1), \quad r_{11} + r_2 \leq \max(1, \alpha).$$

In the following proposition, we characterize the DMT of the open-access policy, whose proof is omitted due to the lack of space.

*Proposition 2:* The diversity order of the system with open-access is given by

$$d_{out}(r_{11}, r_{22}, \alpha) = \min(d_a, d_b, d_c),$$

where

$$d_a(r_{11}) = (1 - r_{11})^+, \quad d_b(r_{22}, \alpha) = (1 - r_{22})^+ + (\alpha - r_{22})^+$$
$$d_c(r_\Sigma, \alpha) = 2(1 - 2r_\Sigma)^+ + (\alpha - 2r_\Sigma)^+$$

with $r_\Sigma = r_{11} + r_{22}$.

In the orthogonal access model, the macro-cell and femto-cell users are allocated orthogonal channel access. The optimized (with respect to $\alpha$) degrees of freedom is then given as

$$r_{\text{orth}} = \frac{\alpha}{1 + \alpha}, \quad (3)$$

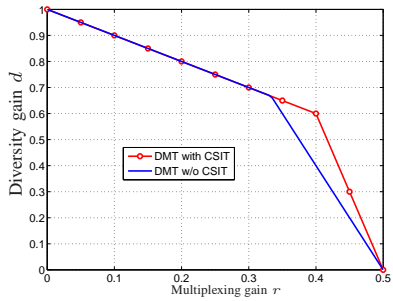

Fig. 2. DMT of the interference channel with and without CSIT.

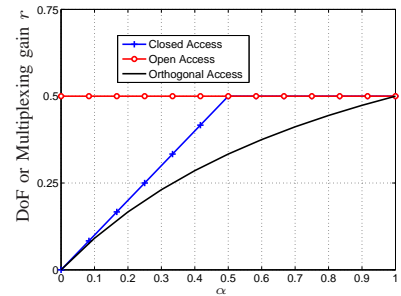

Fig. 4. DoF for open-access, closed-access and orthogonal access policies.

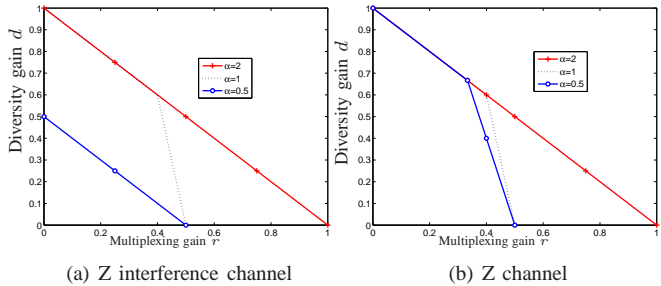

(a) Z interference channel  (b) Z channel

Fig. 3. DMT of the Z channel and the Z interference channel

while the DMT with orthogonal access reduces to that of a single-user system with reduced maximal multiplexing gain $r_{\text{orth}}$.

## V. ILLUSTRATION

In Fig. 2, the DMT for the interference channel with and without CSIT is illustrated for $r = r_1 = r_2$. We observe that up to a multiplexing gain of $r = 1/3$, there is no gain in having CSIT available at the transmitter. The reason for that is that up to $r = 1/3$, the dominant error event is that one of the users is in error. Thus, the DMT is determined by the single-user performance, where the availability of CSIT has no additional gain. Beyond that point, the dominant error event is that all users are in error in contrast to that one of the users is in error. Thus, CSIT becomes valueable and provides some additional diversity gain in comparison to the case without CSIT. In Fig. 3(a) and 3(b), the DMTs are depicted for the Z interference channel and Z channel with $r = r_1 = r_2$, respectively. The DMT for channels are identical for $\alpha = 1$ and $\alpha = 2$, since the link between the macro-cell user and the BS does not represent the bottleneck of the system. However, for $\alpha = 1/2$ the link between macro-cell user and BS becomes the bottleneck of the system. As a consequence, the DMT for the Z interference channel (or closed-access protocol) is strongly dominated by this link, while the DMT for the Z channel (open-access protocol) is only slightly effected.

The DoF (or multiplexing gain $r$) is shown for the open-access, closed-access and orthogonal access policies in Fig. 4 for the asymptotic SNR $\to \infty$ case with $r = r_1 = r_2$. We observe that the closed-access policy achieves the same DoF as the open-access policy as long as $\alpha \geq 0.5$. However, if the link quality between the macro-cell user and BS is further reduced (i.e. $\alpha < 0.5$), the DoF for the closed-access policy is reduced significantly, while the DoF for the open-access policy is unaffected. The orthogonal access policy achieves the same DoF as the closed-access policy only, if $\alpha$ is close to zero or one.

## VI. SUMMARY

In this work, we analyzed the diversity-multiplexing tradeoff for interference networks. It was shown that the DMT of interference networks reduces to the DMT of multiple access channel if the transmitter is not aware of the channel gains. The situation changes if channel state information is available. We also showed that rate-splitting can be ignored for asymptotic analysis, however, it has to be considered for finite SNR. Lastly, we investigated the DoF and the DMT of interference networks with channel knowledge in the context of femto-cells.